\documentclass[11pt,letter,iop]{emulateapj}
\usepackage{amsmath}
\usepackage{natbib}
\usepackage{graphicx}
\usepackage{rotating}
\usepackage{color}

\begin{document}

\title[Being WISE I]{Being WISE I: Validating Stellar Population Models and M$_\star$/L ratios at 3.4 and 4.6$\mu\lowercase{m}$}
\author{Mark A. Norris$^{1}$, Sharon Meidt$^{1}$, Glenn Van de Ven$^{1}$, Eva Schinnerer$^{1}$, Brent Groves$^{1}$ \& Miguel Querejeta$^{1}$}
\altaffiltext{1}{Max Planck Institut f\"{u}r Astronomie, K\"{o}nigstuhl 17, D-69117, Heidelberg, Germany }
\email{norris@mpia.de}
\date{\today}



\def\Re{{R$_{\rm e}$}}
\def\etal{{\it et al. }}

\begin{abstract}
Using data from the WISE mission, we have measured near infra-red (NIR) photometry of a diverse sample of dust-free 
stellar systems (globular clusters, dwarf and giant early-type galaxies) which have metallicities that span the range --2.2 $<$ [Fe/H] 
(dex) $<$ 0.3. This dramatically increases the sample size and broadens the metallicity regime over which the 3.4 (W1) and 
4.6$\mu$m (W2) photometry of stellar populations have been examined.

We find that the W1 - W2 colors of intermediate and old ($>$ 2~Gyr) stellar populations are insensitive to the age of the 
stellar population, but that the W1 - W2 colors become bluer with increasing metallicity, a trend not well reproduced 
by most stellar population synthesis (SPS) models. In common with previous studies, we attribute this behavior to the increasing 
strength of the CO absorption feature located in the 4.6$\mu$m bandpass with metallicity. 

Having used our sample to validate the efficacy of some of the SPS models, we use these models to derive stellar mass-to-light ratios 
in the W1 and W2 bands. Utilizing observational data from the SAURON and ATLAS3D surveys, we demonstrate that these 
bands provide extremely simple, yet robust stellar mass tracers for dust free older stellar populations that are freed from many of the 
uncertainties common among optical estimators.

\end{abstract}

\section{Introduction}

Stellar population synthesis models are an important tool to investigate the formation and
evolution of galaxies. By comparison of SPS models with observed galaxy spectral energy 
distributions (SEDs) it is possible to infer the mass of stars, gas, and dust, as well as the age 
and chemical enrichment of the underlying stellar populations 
\citep[see e.g. ][]{Walcher11_Review,Conroy13_Review}. 

Much recent work has focussed on the development of SPS models that accurately describe the 
behavior of galaxies in the near-to-mid infrared \citep[e.g.][]{daCunha08,Marigo08,GALEVI,FSPS,
Bressan12,Bressan13}. In part because the near infrared bands are thought to be a more accurate 
proxy for total stellar mass than optical regions, due to the reduced importance of dust attenuation or 
short-lived but luminous young stars in IR bands. Importantly, however, dust $\textit{emission}$ can 
significantly effect the NIR colors of stellar systems (\citealt{Meidt12a,SAURONXX}; Querejeta et al. 2014) 
and must therefore be carefully taken into account when using NIR photometry as a stellar mass tracer.
Yet, despite the rapid increase in theoretical predictions for the behavior of galaxy SEDs in 
the near and mid IR, to date only a limited number of studies have calibrated these predictions to 
observations of real stellar systems. 

\citet{Spitler08} compared the predictions of a range of Single Stellar Population (SSP) models to 
the optical - $\textit{Spitzer}$ IRAC [3.6] colors of a sample of globular clusters (GCs) from the 
Sombrero and Centaurus A galaxies. They found generally good agreement in the colors, though 
with some evidence that the models could be under predicting the [3.6] flux at lower metallicities.

\cite{SAURONXX} then investigated the behavior of the $\textit{Spitzer}$ IRAC [3.6] - [4.5] color 
of the SAURON sample of early-type galaxies \citep{SAURONII}. They concluded that the galaxies displayed 
a color trend with metallicity over the limited metallicity range that they probed ([Fe/H] $>$ -0.5). 
The sense of this trend was reversed relative to that observed in the optical (i.e. that in this case 
increasing metallicity led to bluer colors), and was not well reproduced by SPS models except
for ones which included the influence of a strong CO absorption band in the [4.5] filter 
whose strength is temperature dependent.

Subsequently \citet{Barmby12} compared SSP models to V band and $\textit{Spitzer}$ IRAC photometry 
of bright globular clusters (GCs) from M31 also finding that the models examined were systematically redder 
than the data at higher metallicity. 

Each of these previous studies suffered from either limited wavelength coverage in the NIR \citep{Spitler08},
restricted metallicity range of the sample studied \citep{SAURONXX}, or limited statistical power due to 
a small sample size \citep{Barmby12}, or some combination of all three.

In light of these results and the recent explosion in the amount of galaxy photometry available in the 
NIR, provided by both large surveys using the Spitzer Space Telescope \cite[e.g. S$^{4}$G;][]{S4G} and the all sky 
coverage of the WISE mission \citep{Wright10}, we have revisited the behavior of stellar populations in 
the NIR. We do this by analyzing a comprehensive sample of globular clusters and early-type galaxies in 
order to probe the color-metallicity behavior over a very wide metallicity range. We focus on 
GCs and early-type galaxies because these systems are essentially dust-free \citep[e.g.][]{BArmby09,HCVSXVIII}, 
and as described previously dust emission can significantly alter the NIR colors of stellar systems.

Our work is further motivated by recent work by \citet{Meidt14a} who demonstrated that the [3.6] - [4.5] 
color can be used to improve the estimation of 3.6$\mu$m stellar mass-to-light ratio. To do this they 
empirically derived a relation between the J-K and [3.6] - [4.5] colors of giant stars, thereby avoiding the 
uncertainties present in the SPS models due to incorrect molecular line opacities of the template stars in
the NIR. In doing this they provided a significantly improved stellar mass estimator than available at 
shorter wavelengths. However, the empirical relation needs to be verified for complex stellar populations 
over the age and metallicity ranges where it is likely to be used.

\section{Data Analysis}

For each of our samples being considered (see Section \ref{Sec:Sample}), except for the globular
clusters of M31 which are essentially unresolved in WISE imaging, we downloaded and analysed the 
full (i.e. not thumbnail) ALLWISE\footnote{http://wise2.ipac.caltech.edu/docs/release/allwise} Atlas images. 
These images have been pipeline processed as described in \cite{Wright10} and \cite{Mainzer11} to produce 
co-added image tiles with uniform zeropoint that are spatially co-aligned in all four WISE bands. They are 
the result of co-adding all available WISE imaging, including NEOWISE imaging not previously included in 
earlier data releases, hence they represent the deepest WISE imaging available to date.

We choose to analyse WISE imaging (as opposed to Spitzer IRAC imaging) of our old stellar systems 
for a number of reasons; 1) WISE has consistent all-sky coverage ensuring large sample sizes. 2)
The survey is relatively deep, with at least S/N = 5 for W1 = 16.9 and W2 = 16.0 mag, approximately 
twice as deep as 2MASS. 3) The large (1.56$^\circ$$\times$1.56$^\circ$) field-of-view of the processed atlas images 
reduces sky subtraction problems for large apparent size objects such as Milky Way (MW) GCs and nearby massive 
early-type galaxies which often have half-light radii $>$ 3\arcmin. 4) The resolution of the W1 and W2 
bands while low by optical standards ($\sim$8.5\arcsec\, sampled with 1.375\arcsec\, pixels) is more 
than adequate to study nearby GCs and galaxies.

\subsection{Stochastic Effects}
\label{Sec:stochastic}

When studying the integrated properties of lower mass stellar populations such as globular clusters it is 
necessary to consider to what degree measured quantities depend on the stochastic population of rare but 
luminous stellar phases, such as the red giant or asymptotic giant branches. The random appearance of
a small number of stars on these evolved branches can lead to dramatic changes in both the total luminosity
of the stellar population, and its color \citep[see e.g.][]{Fouesneau10,Popescu10}. Especially in the optical 
where the colors of extreme red giant branch stars and the turnoff stars can differ by as much as a magnitude.

In contrast to the usual approach of theoretically determining the minimum stellar mass required to
reduce stochastic population effects to manageable levels (see e.g. \citealt{Barmby12}) we choose to use
an empirical procedure to determine the necessary stellar mass to limit the effect of stochastic population of the 
evolved phases on the measured color. To do this we have measured the W1-W2 color within the half-light-radius 
for all suitable MW GCs with M$_{\rm V}$ $<$ --7 (corresponding to total luminosities sampled 
of 0.75 mag less due to our approach of only sampling to one effective radius).

\begin{figure} 
   \centering
   \begin{turn}{0}
   \includegraphics[scale=1.05]{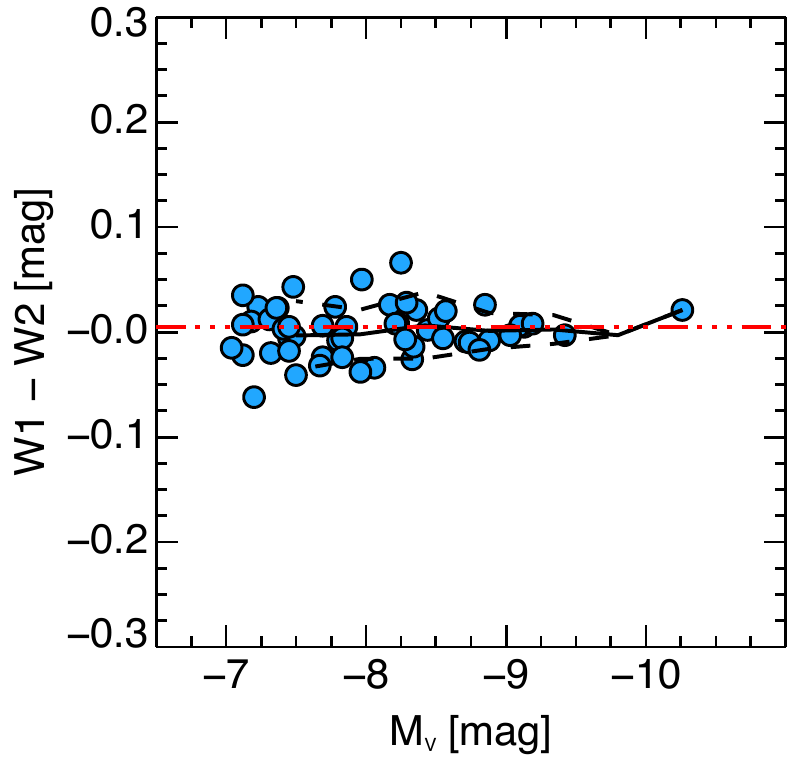}
   \end{turn} 
   \caption{W1-W2 color measured within R$_{\rm e}$ for Milky Way globular clusters as a function of total V-band 
   absolute magnitude. The solid black line shows the running average and the dashed lines the scatter.
   The red triple-dot-dashed line show the median W1 - W2 color of all GCs where the error in the W1 - W2 color is 
    less than 0.05 mag.
       }
   \label{fig:stochastic}
\end{figure}

As shown in Figure \ref{fig:stochastic} the magnitude dependence of the scatter of the measured W1 - W2 colors 
is remarkably small, with very little dependence on total magnitude of the cluster other than that expected from 
purely photometric errors. This result is unexpected given that the scatter in this plot is composed of three effects; 
1) photometric uncertainties, 2) the effect of the metallicity dependence of the color, and 3) the effect of stochastically 
populating the evolved branches. Therefore, given the observed low scatter, the known photometric errors (of around
0.01 mag for Milky Way GCs), and expected color trend the effect of stochastic sampling must be relatively small.

\begin{figure} 
   \centering
   \begin{turn}{0}
   \includegraphics[scale=0.55]{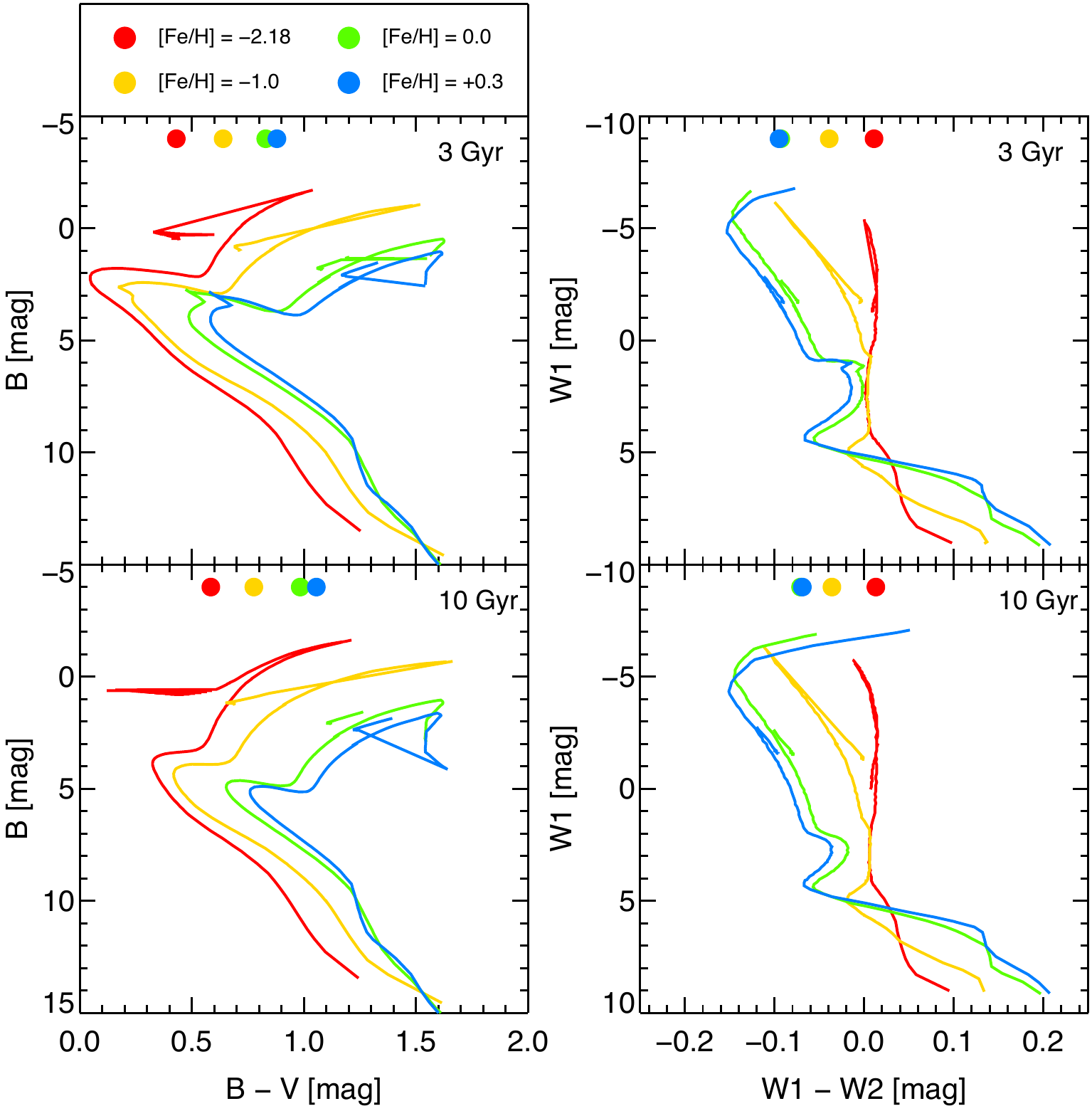}
   \end{turn} 
   \caption{Stellar isochrones for single stellar populations from the \citet{Bressan12,Bressan13} models for both B - V (left panels) 
   and W1 - W2 (right panels). The upper panels show the isochrones for 3 Gyr stellar populations, and the lower panels are 10 Gyr 
   SSPs. The colors represent different metallicities as indicated in the legend at the top.The colored circles at the top in each panel 
   show the total integrated color of each SSP assuming a Chabrier IMF and a fully populated isochrone. The integrated color is 
   dominated by the color of stars on or near the main sequence turnoff, but only in the optical case do red giant branch stars have 
   colors that differ considerably from this average color. Note also that the metallicity dependence of the color is reversed for the 
   NIR colors with more metal rich isochrones becoming bluer in the NIR case. }
   \label{fig:isochrones}
\end{figure}

We investigate this point more closely in Figure \ref{fig:isochrones}. This figure shows stellar isochrones for both
optical and NIR filters for 8 SSPs produced by the PARSEC v1.1 models of \citet{Bressan12,Bressan13}. The 
colored circles at the top of the plot show the total integrated color of the SSPs for a Chabrier lognormal IMF.
From this plot it is clear that the colors of the main sequence turnoff (which dominates the overall color) and the
extreme RGB differ by at most 0.1 mag in the NIR, but by up to 1.1 mag in the optical, hence population of the
extreme RGB does little to change the color of the population for NIR bands.

\subsection{Stellar Population Samples}
\label{Sec:Sample}

\subsubsection{Milky Way and Magellanic Cloud Globular Clusters}

We selected an initial sample of MW GCs drawn from the \citealt{Harris96} (2010 edition) catalog, supplemented with a 
sample of Large and Small Magellanic Cloud GCs drawn from \citet{McLaughlin05}. The sample was created by requiring 
the GCs to 1) have a measured age and metallicity, 2) to be older than 2 Gyr, 3) be more luminous than M$_{\rm V}$ = --7.00, 
and 4) yet not saturated in the WISE imaging, and finally 5) to have half-light radii (R$_{\rm e} >$ 10\arcsec) that are well 
resolved by the W1 and W2 photometry. The third requirement ensures that the GCs are sufficiently massive that the effect 
of stochastic variation in the population of their red giant and horizontal branches is negligible (see Section \ref{Sec:stochastic}). 
The fourth requirement is more subtle, as the ALLWISE combination algorithm tends to hide saturation in the final images 
in all but the most extreme cases. As saturation is more common in the W1 band this can lead to spurious color trends. 
Only by checking the single epoch WISE exposures manually was it possible to be certain that the atlas images 
were unaffected by saturation. 

By applying these requirements (and a subsequent cut on the photometric uncertainty of the W1 - W2 color) we are left with 
a final sample of 66 MW and Magellanic Cloud GCs. Unfortunately because of the saturation problems, which preferentially 
effect closer and higher metallicity MW GCs, the sample has relatively few GCs with metallicity [Fe/H] $>$ --1.

Each cluster was analysed in the following manner:

\begin{enumerate}

\item The ALLWISE Atlas science and uncertainty images for each cluster were downloaded. 

\item Using the literature values of GC center, position angle, ellipticity (generally negligible for GCs) and R$_{\rm e}$, 
an aperture containing half the GC light was defined and the flux summed within this region. The value of 
R$_{\rm e}$ was determined from optical data and was therefore scaled by a factor of 0.71 to account for the fact that 
the half-light radius is smaller in IR bands \citep{SAURONXIX}. Summing the flux within R$_{\rm e}$ ensures that the 
flux of the GC dominates and the influence of background sources is negligible. It also ensures that we can measure 
the flux within a single aperture, rather than have to carry out the uncertain task of identifying and measuring individual 
cluster stars at large radii where the field contamination becomes significant.

\item SExtractor \citep{Bertin_sextractor} was used to determine a background map, after masking the GC, 
and using a very large BACK$\_$SIZE value of 512 pixels to ensure that the influence of foreground and 
background sources was reduced.

\item The ``sky" background was then subtracted from the total GC flux in each band, the magnitudes
and their uncertainties calculated and the standard conversions to the WISE magnitude system applied 
\citep{Wright10}.

\end{enumerate}

\subsubsection{M31 Globular Clusters}
\label{sec:m31gcs}

To augment the MW GC sample we also make use of WISE photometry of spectroscopically confirmed 
M31 GCs drawn from the Revised Bologna Catalog of M31 GCs \citep[RBC;][]{BolognaM31GCs}. To the RBC
we add additional age and metallicity measurements from the literature \citep{Ma09,Wang10,Caldwell11,Cezario13}
to produce the largest possible catalog of spectroscopically confirmed M31 GCs with stellar population
measurements.

We apply similar requirements to those used for the MW GC system. We also require spectroscopic
confirmation of their nature, as the RBC catalog suffers from significant contamination from foreground and background 
objects (see e.g. \citealt{Huxor14}). Saturation of the WISE imaging is not an issue for M31 GCs due to the increased 
distance relative to the MW GCs, however, we do visually inspect each cluster to reject those cases where 
the photometry is likely compromised by bright nearby sources. 

After this selection we are left with a sample of 61 M31 GCs, with metallicities that range from [Fe/H] of --1.9 to --0.1. 
The photometry for the M31 GC sample is taken directly from the catalog measurements for the ALLWISE data release. 
In this case we do apply an extinction correction, as some of the GCs are observed in (and even behind) the disk of 
M31. Where available we make use of the \citet{Caldwell11} estimations of E(B-V), because these include the foreground 
MW extinction and the effects of the internal M31 extinction. Where our GCs lack E(B-V) measurements we use the median
value of the E(B-V) for the M31 GCs as determined by \citet{Caldwell11}. We then convert the E(B-V) into
extinctions in the W1 and W2 bands using the conversions quoted in \cite{Yuan13}.

\subsubsection{Early-Type Galaxies}
\label{sec:etgs}

In addition to the GC samples we also examine the WISE colors of early-type galaxies (ETGs). Our main ETG 
sample is that of the SAURON survey \citep{SAURONII}. This contains of 48 nearby S0 and E galaxies observed 
with the SAURON spectrograph on the WHT telescope \citep{SAURON1}. The NIR colors of these galaxies 
have already been examined by \cite{SAURONXX}, who used Spitzer IRAC imaging of the sample galaxies 
to measure their [3.6] - [4.5] colors, finding that galaxies became increasingly blue with increasing 
metallicity, in marked contrast to the behavior for most other colors. The principle limitation of the SAURON 
sample is that the metallicity range is quite restricted, with most galaxies having --0.5 $<$ [Fe/H] $<$ 0. 
Therefore we have extended the metallicity range probed by the galaxy sample by adding additional dwarf and giant 
early type galaxies to the sample. 

In order to be considered for inclusion in our extended galaxy sample several properties had to be available in 
the literature. These comprised suitable photometric values for effective radius, position angle, and ellipticity as 
well as spectroscopically determined age and metallicity, (ideally weighted to be within R$_{\rm e}$). Such 
information was compiled from the literature sources listed below, occasionally supplemented by position angles 
and ellipticities provided by either Hyperleda \citep{Hyperleda} or the SDSS data release 10 \citep{SDSSDR10}.

We added several lower mass dE and dS0 galaxies from the work of \cite{Rys13,Rys14}. These galaxies were
observed using the SAURON spectrograph similar to the main SAURON sample, with their stellar populations 
(Rys et al. in prep) also derived in a manner consistent with the rest of the SAURON sample.

We also added additional dwarf and giant early-type galaxies using derived properties provided in the papers of 
\citet{Michielsen08,Koleva11,Forbes11}.

Finally, we added higher-mass early-type galaxies from the study of \cite{Denicolo05}. In this case we converted
their measured central metallicity to an approximate half-light weighted metallicity by subtracting the mean
offset between the R$_{\rm e}$ and R$_{\rm e}$/8 derived metallicities of the SAURON sample. 

Where necessary (for example with the SAURON sample) we convert their measured [Z/H] metallicity estimates
to [Fe/H] using the relation that [Z/H] = [Fe/H] + 0.94$\times$[$\alpha$/Fe] \citep{Thomas03}.

\begin{figure} 
   \centering
   \begin{turn}{0}
   \includegraphics[scale=0.575]{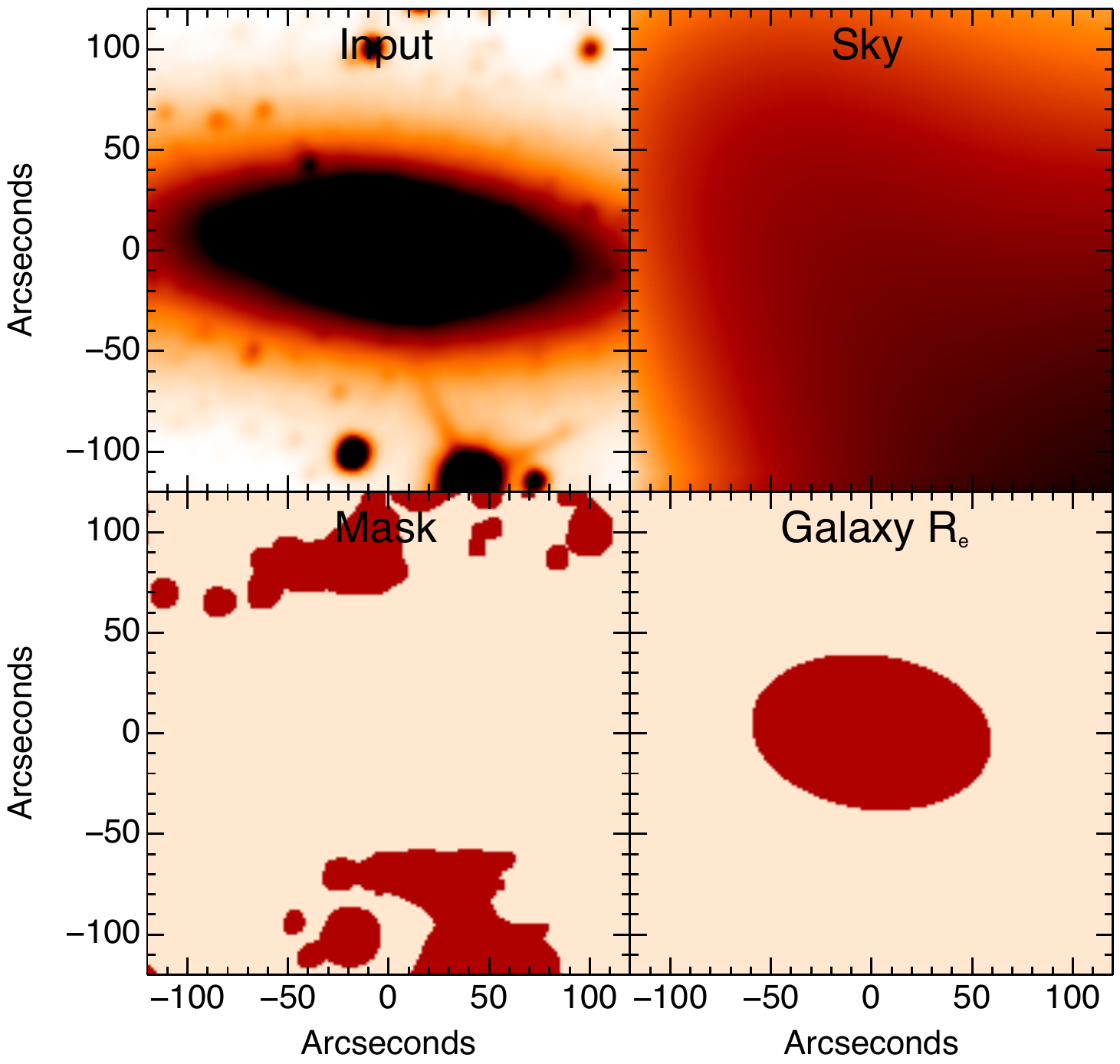}
   \end{turn} 
   \caption{Photometric analysis of the early-type galaxy NGC~~1023. $\bf{Top\,Left:}$
   Zoomed region of the input WISE W1 image. $\bf{Top\,Right:}$ The SExtractor derived sky map (with stretch altered
   to enhance spatial variability).  No obvious sign of influence
   from the galaxy itself can be seen. $\bf{Bottom\,Left:}$  The
   SExtractor derived masked regions.  $\bf{Bottom\,Right:}$ The elliptical half-light isophote.}
   \label{fig:photometry}
\end{figure}

We analyse the full galaxy sample in the same way as the MW GCs, i.e. we measure the WISE photometry within 
ellipses set by the observed ellipticity, position angle and R$_{\rm e}$ of the galaxies. Figure \ref{fig:photometry} 
displays an example of this procedure. Likewise we restrict the sample selected for study using the same limitations, 
for example we only study those objects which have R$_{\rm e} >$ 10\arcsec\, to ensure that the half-light radius is 
adequately resolved in the WISE imaging. We also limit our analysis to galaxies which have age $>$ 2 Gyr, to 
ensure that the effect of poorly understood evolved stellar phases such as thermally pulsating 
asymptotic giant branch (TP-AGB) is significantly reduced \citep[see e.g][]{Maraston05}.

\section{Results}

\subsection{W1-W2 colors of Old Stellar Populations}
\label{sec:color_met}

\begin{figure*} 
   \centering
   \begin{turn}{0}
   \includegraphics[scale=1.15]{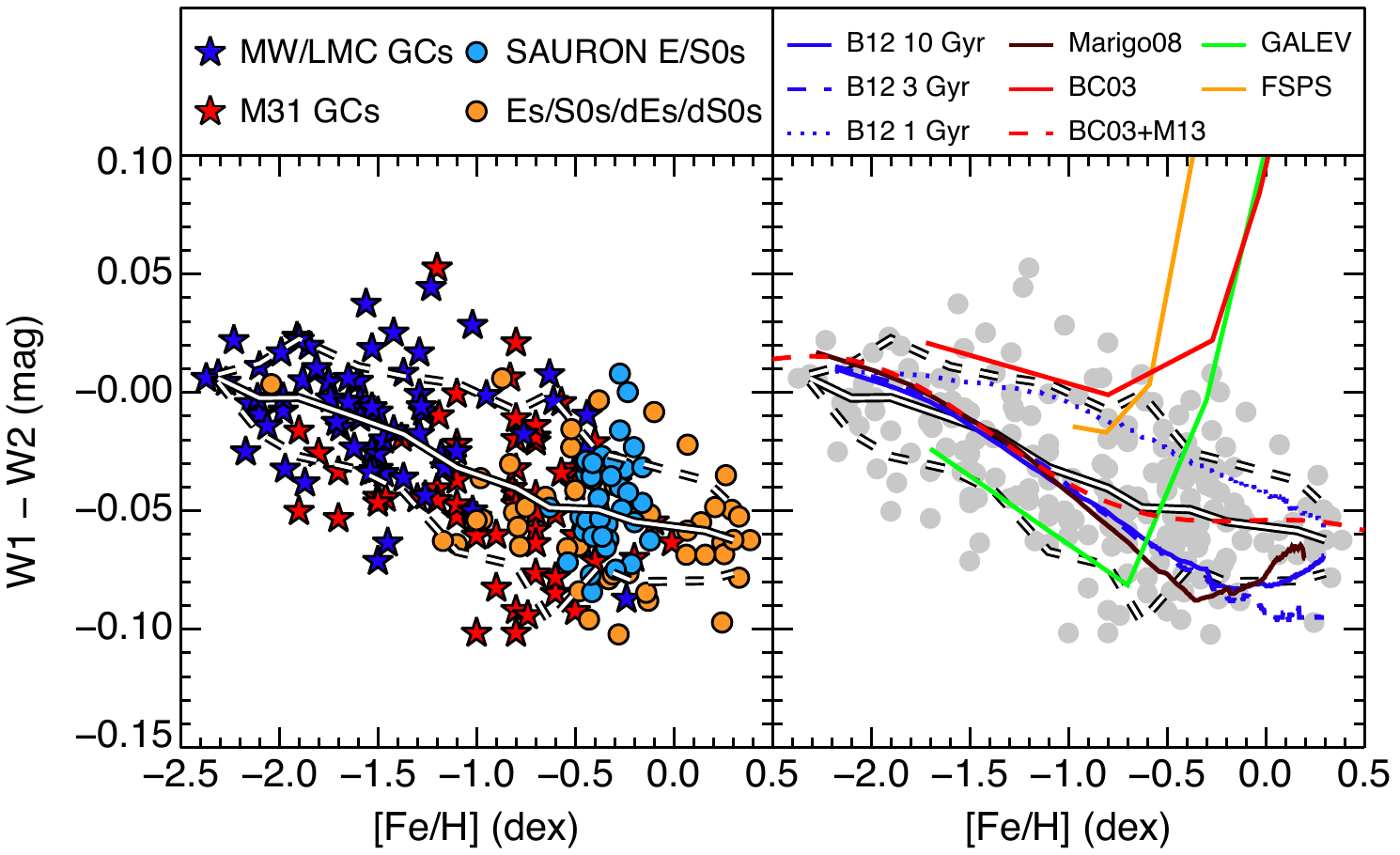}
   \end{turn} 
   \caption{$\bf{Left\,Panel:}$ The W1 - W2 colors of our sample of dust-free stellar systems plotted against their metallicity.
   The blue stars are MW and Magellanic Cloud GCs, the red stars are M31 GCs, the cyan circles are the SAURON 
   sample galaxies, and the orange circles are the remaining dwarf and giant early-type galaxies (See Section 
   \label{Sec:Sample} for details). The black-and-white solid line shows the running average of the whole sample, as 
   determined in 0.25 dex bins, the black-and-white dashed lines are the 1 $\sigma$ scatter on the running average.
   $\bf{Right\,Panel:}$ The grey circles show the entire sample, now undifferentiated by stellar system type or source.
   The solid and dashed black-and-white lines show the same running mean and 1 $\sigma$ scatter from the left panel.
   The blue lines show predictions of the \citet{Bressan12,Bressan13} SPS model for stellar populations with ages of 
   1 Gyr (dotted line), 3 Gyr (dashed line), and 10 Gyr (solid line). All remaining models are for 10Gyr SSPs: 
   The brown solid line shows the \citet{Marigo08} model with the \citet{Girardi10} TP-AGB tracks. The solid red line 
   shows the \citet{BruzualCharlot} prediction, while the dashed red line shows \citet{BruzualCharlot} prediction after 
   empirical correction for the effects of the 4.5$\mu$m CO feature by \citet{Meidt14a}. The solid green line shows 
   the prediction from the GALEV model of \citet{GALEVI}. Finally the solid orange line shows the prediction from the 
   FSPS models of \citet{Conroy09} and \citet{Conroy10}.}
   \label{fig:FeH_color}
\end{figure*}

The left panel of Figure \ref{fig:FeH_color} displays the W1 - W2 colors of our sample of old stellar systems plotted 
against their metallicity. The solid and dashed black and white lines are the running average of the whole sample and 
the 1 $\sigma$ scatter, as measured in 0.25 dex bins. It is clear from this plot that there is a highly significant trend, with 
the stellar populations becoming bluer with increasing metallicity, as was found previously by \cite{SAURONXX,Meidt14a}
for the IRAC [3.6] - [4.5] $\mu$m color. The running average of the points is well reproduced by a linear relation 
between metallicity and W1-W2 color of the following form:

\begin{eqnarray}
W1- W2 &=& -0.026\times[Fe/H](dex) - 0.058
\end{eqnarray}

with the scatter about the relation being 0.026 magnitudes.

The right panel of Figure \ref{fig:FeH_color} displays the same data, but overlaid with several current SPS model predictions.
The SPS models fall into two groups; those that attempt to correct for the CO absorption in the W2 band (the PARSEC v1.1
models of \citealt{Bressan12,Bressan13}, the earlier Padova group models of \citealt{Marigo08} using the \citealt{Girardi10} TP-AGB 
tracks, and the \citealt{Meidt14a} corrected \citealt{BruzualCharlot} models), and those that do not (the \citealt{BruzualCharlot},
the GALEV models of \citealt{GALEVI}, and the FSPS model of \citealt{Conroy09,Conroy10}).

It is clear that the majority of the models, including those most commonly used to derive stellar population properties and stellar
masses dramatically fail to reproduce the W1 - W2 colors of stellar populations with near solar metallicity. Only those models that 
include the effect of the increasing CO absorption strength at 4.5$\mu$m successfully reproduce the observed trend. Both the 
Padova group models and the empirical model fit the data reasonably, with the empirical model of \citet{Meidt14a} being slightly 
more consistent with the data overall. 

Note that, based on the behavior of the \citet{Bressan12,Bressan13} models, the W1 - W2 color is almost insensitive 
to age for stellar populations $>$ 2 Gyr. We therefore do not expect that age contributes significantly to the observed 
scatter here, which is large even for relatively bright stellar systems.  Such large observational scatter unfortunately prevents us
currently using W1-W2 color as an accurate metallicity indicator. Still, we note that improvements in data quality may
allow the use of the W1-W2 color may act as a useful a prior on metallicity, e.g. to help break the well-known age-metallicity 
relation in the optical.

\subsection{The Absolute Zeropoint in the W1 \& W2 Bands}

In order to confirm the zero point of the models in the WISE filter system we can use the models to predict 
the absolute magnitude of the Sun in the WISE bands.

The \citet{Bressan12,Bressan13} PARSEC v1.1 isochrones give the absolute magnitude of a solar mass star with 
a solar abundance and age 4.6 Gyr as W1 = 3.24 and W2 = 3.26. These values are remarkably close to the values 
of 3.24 and 3.27 as determined by \citet{Jarrett13} for W1 and W2 based on the WISE relative system response and 
the SED of the Sun. The \citet{Marigo08} models likewise predict solar magnitudes very close to the observed values; 
3.22 and 3.24 magnitudes, respectively.

In contrast, while the remaining models (BC03, FSPS, and GALEV) do approximately give the correct value for the
magnitude of the sun in the W1 band, as expected from their predictions of the W1 - W2 color (see Section \ref{sec:color_met}), 
they substantially underpredict the absolute magnitude of the Sun in the W2 band.

As the PARSEC models correctly predict both the absolute zero point, and the color behavior of the W1 and W2 
filters we therefore choose to examine further whether these models and bands can be used to produce a widely 
applicable stellar mass estimator.

\subsection{Mass-to-light ratios at 3.4 and 4.6$\mu$m }
\label{sec:m_l}

Having demonstrated that some modern SPS models are capable of reproducing the observed WISE
W1 and W2 photometry of dust-free stellar populations we now examine what these models predict for the
mass-to-light ratio behavior of simple stellar populations.

\begin{figure*} 
   \centering
   \begin{turn}{0}
   \includegraphics[scale=1.00]{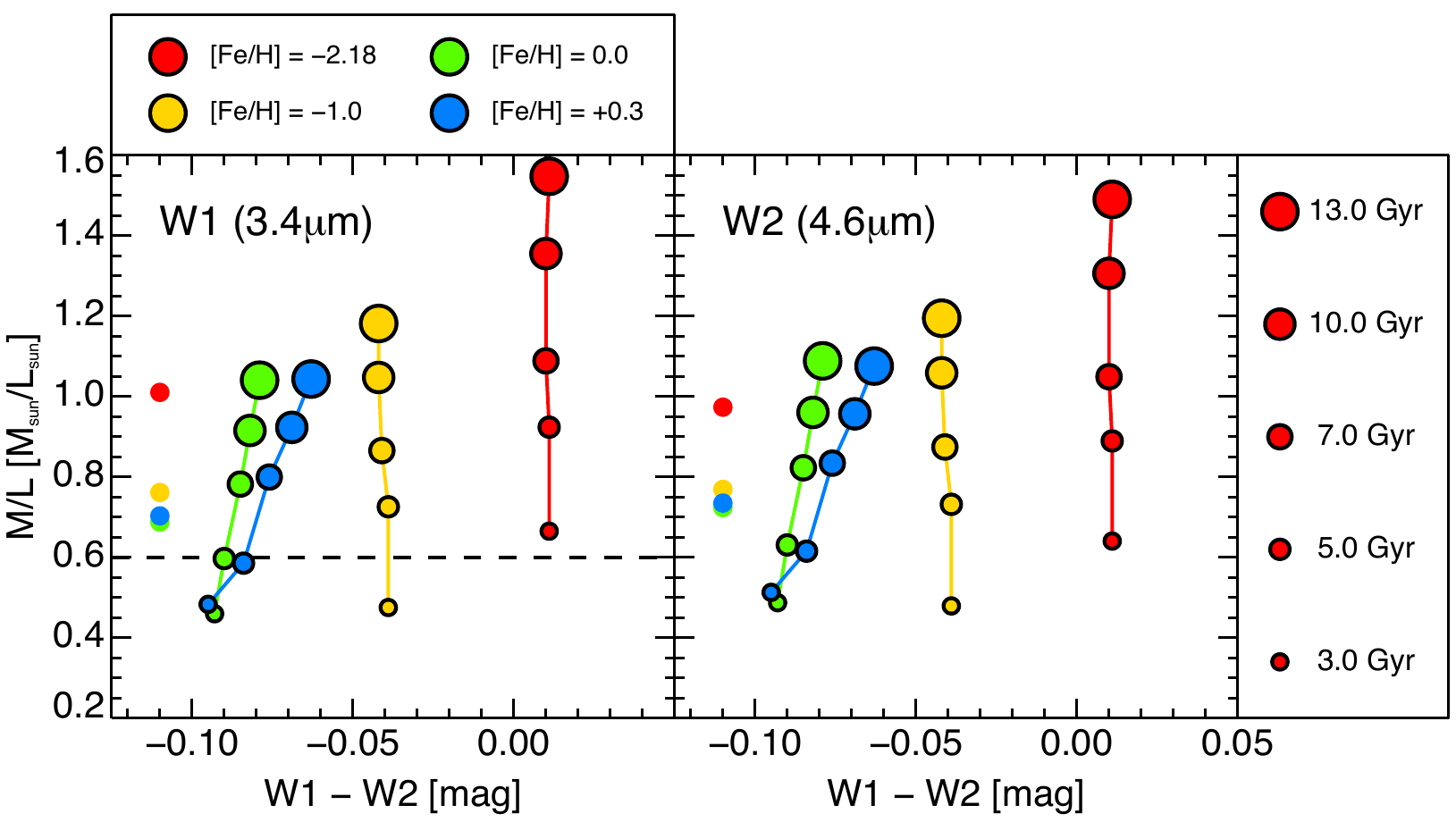}
   \end{turn} 
   \caption{The W1 - W2 color vs 3.4 and 4.6$\mu$m mass-to-light ratios derived from the PARSEC v1.1 models of
   \cite{Bressan12,Bressan13} as described in Section \ref{sec:m_l}. Single stellar populations for combinations of 
   four metallicities ranging from --2.18 to +0.3, and ages from 3 to 13 Gyr are displayed. The colored dots to the
   left of each panel without black outline display the average M$_\star$/L for SSPs with age between 3 and 10 Gyr 
   (M$_\star$/L$_{\rm W1}$ = 1.01, 0.76, 0.69 and 0.70 for the 4 metallicities plotted). The 
   dashed line in the left panel displays the M$_\star$/L of 0.6 suggested by \citet{Meidt14a} as a suitable fixed M$_\star$/L for
   old stellar populations, based on \citet{BruzualCharlot} models with exponentially declining star formation histories.
   As noted by \citet{Meidt14a} the mass-to-light ratio increases by a factor of around 2 between 3 and 10 Gyr.}
   \label{fig:m_l}
\end{figure*}

To do this we make use of the PARSEC v1.1 models \citep{Bressan12,Bressan13} for a Chabrier lognormal IMF
and single burst models (i.e. pure SSPs) which we convert to mass-to-light ratios assuming that the mass remaining 
in living stars and remnants follows the tracks presented in \citet{Into13} for a Kroupa IMF (i.e. $\sim$30$\%$ of the 
stellar mass is returned to the ISM within 12 Gyr). Figure \ref{fig:m_l} shows the result of this procedure for a range of 4 
metallicities (from [Fe/H] = --2.18 to +0.3) and eight ages from 0.5 Gyr to 10 Gyr. It is important to note that because 
the current v1.1 of the PARSEC models do not include the effects of TP-AGB stars, the predictions for ages $<$ 3 
Gyr are likely to be significantly in error. However, comparison with the \citet{Marigo08} models, which are an earlier
iteration of the Padova models which do include the effects of TP-AGB, shows that for ages $>$ 3 Gyr the predicted
M$_\star$/L ratios seem to be robust. 

Examining Figure \ref{fig:m_l} it is immediately obvious that in common with the conclusions of \cite{Meidt14a} we find
that for metallicities displayed by modern galaxies (i.e. [Fe/H] $>$ --1) the M$_\star$/L ratios are essentially insensitive to
metallicity (and hence color). It is also notable that in the age range of 3 to 10 Gyr, the M$_\star$/L ratio increases by around
a factor of two, meaning that applying a fixed M$_\star$/L ratio in the middle of the range leads to errors on the derived stellar 
mass of only 0.10 dex. This is also in good agreement with the analysis of \citet{Meidt14a}, who using their modified 
\citet{BruzualCharlot} models found that for exponentially declining star formation histories (which tend to reduce the 
overall M$_\star$/L when compared to single burst histories) it was possible to use a single M$_\star$/L of 0.6 to derive stellar masses 
from IRAC [3.6] photometry with 0.10 dex uncertainties.

As expected, the derived M$_\star$/L ratios for the 3.4 and 4.6$\mu$m bands are very similar, and that therefore both have the
potential to be useful stellar mass indicators. In general the 3.4$\mu$m band is likely to be preferred due to the higher
S/N of the WISE imaging in this band. However, in certain cases, the 4.6$\mu$m flux may be a more robust tracer of old 
stellar light, e.g. when the 3.4$\mu$m bandpass also contains strong 3.3$\mu$m PAH emission.  This effect is likely negligible 
for the old stellar populations in this study.

\subsection{Stellar Masses}

In order to check the efficacy of the M$_\star$/L ratios in Section \ref{sec:m_l} we compare WISE derived stellar 
masses to those determined by the ATLAS3D survey \citep[specifically the Salpeter masses from][]{ATLAS3DXX} 
for the 47 galaxies in common with the SAURON survey. We study only the SAURON survey galaxies instead of 
the full ATLAS3D sample because spectroscopically determined stellar population parameters \citep[from ][]{Kuntschner10} 
are only available currently for the SAURON subset. This allows us to examine the improvement in derived stellar 
masses when the additional information provided by the age and metallicity of the galaxies are included. We choose to use the
ATLAS3D Salpeter stellar masses because these masses are computed using the input provided by a
spectroscopically-derived smoothed star formation history. This procedure should therefore reduce the
confusing effect of SFH on derived stellar masses, and provide a more robust stellar mass for comparison to our
WISE derived stellar masses.

To determine the WISE stellar masses our approach is very simple. We take the W1 and W2 luminosity 
of the galaxies measured within R$_{\rm e}$ as described in Section \ref{sec:etgs} and double it to account 
for the flux outside the half-light aperture. To ensure consistency with the ATLAS3D studies we use their 
distance measurements for each galaxy. Once we have the total luminosity of each galaxy we then use two 
different approaches to derive the final stellar mass. The simplest approach is to apply a fixed M$_\star$/L, 
this is chosen arbitrarily to match the zero point of the ATLAS3D masses (once they have been offset by the 
standard +0.25 dex to account for differences between a Chabrier and Salpeter IMF). The second approach 
is to use the measured stellar population parameters of each galaxy to determine the appropriate M$_\star$/L 
through interpolation of the relations shown in Figure \ref{fig:m_l}.

\begin{figure*} 
   \centering
   \begin{turn}{0}
   \includegraphics[scale=1.00]{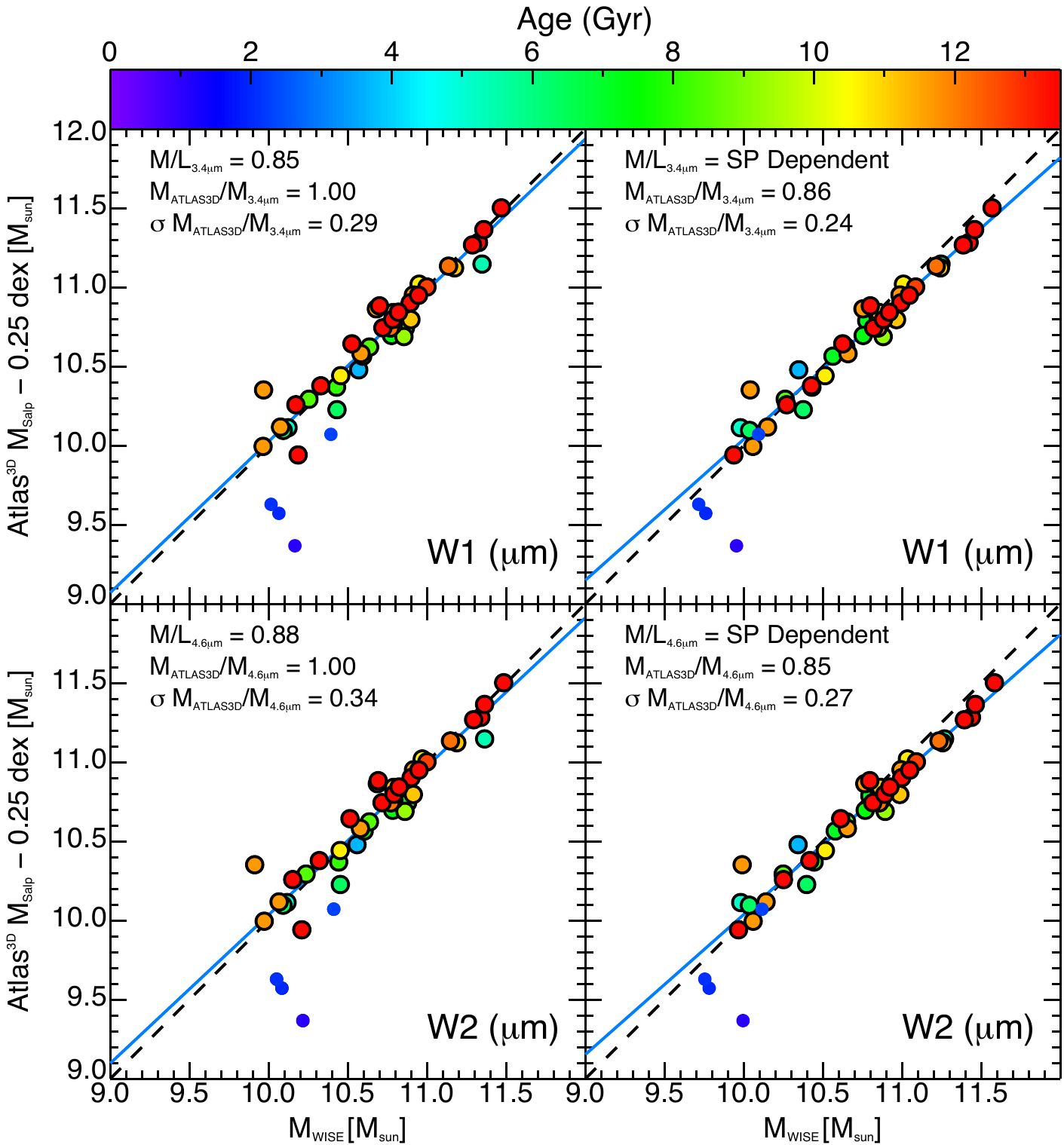}
   \end{turn} 
   \caption{The WISE W1 and W2 derived stellar masses for the SAURON survey \citep{SAURONII}
   galaxies vs the stellar masses derived for the same galaxies by the ATLAS3D survey \citep{ATLAS3DI}.
   In each panel the color of the dots gives the luminosity weighted SSP age as measured by \citet{Kuntschner10}.
   Those galaxies which have measured age within R$_{\rm e}$ of $>$ 3 Gyr have a black outline, those younger
   than 3 Gyr do not. The dashed line in each panel shows the one-to-one relation, while the solid blue line is the 
   best fit linear relation for those galaxies with age $>$ 3 Gyr.
   The stellar masses from the ATLAS3D survey are the Salpeter IMF masses from \citet{ATLAS3DXX}
   scaled by 0.25 dex to account for the difference between the Salpeter IMF and the Chabrier IMF
   assumed for the WISE measurements. The upper panels show the results for the W1 (3.4$\mu$m)
   band and the lower panels for the W2 (4.6$\mu$m) band. The left panels show the result when the
   W1 and W2 mass-to-light ratios are fixed for all galaxies, at a value chosen arbitrarily to match the
   ATLAS3D masses, but consistent with a stellar population with age $\sim$9 Gyr (see Figure \ref{fig:m_l}).
   The right panel show the effect of varying the mass-to-light ratio based on the SAURON measured
   stellar populations \citep[presented in][]{Kuntschner10}. 
   The top right of each panel shows the band used, the method used to determine M$_\star$/L, and both the ratio 
   between ATLAS3D mass and WISE mass and the scatter around this ratio for galaxies older than 3 Gyr. 
    }
   \label{fig:stellar_mass}
\end{figure*}

Figure \ref{fig:stellar_mass} displays the results of these procedures for both the W1 (upper panels) and W2 
(lower panels) bands. From the left panels of the figure, which display the result of using fixed M$_\star$/L ratios, 
it is obvious that the relation is remarkably tight, with most of the significant outliers being known young galaxies 
(the blue dots without black border) where the chosen M$_\star$/L is particularly inappropriate. It should be noted 
that the derived values of M$_\star$/L (0.85 and 0.88 in W1 and W2 respectively) are likely consistent with 0.6 
value derived by \citet{Meidt14a} when it is considered that our M$_\star$/L ratio likely includes some systematic 
offset caused by missing galaxy light due to our method of doubling the flux within half-light apertures, especially 
as the half-light radii are determined in optical bands not directly from the WISE imaging. In fact, if matched apertures 
are used (i.e. the WISE apertures are matched to the ones used by the SAURON survey) the M/L ratios become 
0.67 and 0.70 for the W1 and W2 bands respectively, even more consistent with the 0.6 value derived by 
\citet{Meidt14a}. The observed scatter in the 3.4$\mu$m relation is also remarkably close to the value of 0.1 dex 
predicted by \citet{Meidt14a}, all the more remarkable when it is considered that the errors in the ATLAS3D stellar 
mass determinations must also be considerable.

The right panels display the effect of using M$_\star$/Ls determined using the stellar population parameters,
in this case no offsetting to match the ATLAS3D determinations is done. Several interesting effects
are apparent in these panels: 

\begin{enumerate}

\item The offsets of the younger galaxies from the one-to-one line are significantly reduced, due to their now 
having more appropriate (and lower) M$_\star$/L than in the fixed M$_\star$/L case. 

\item Secondly, there is a significant offset between the ATLAS3D and WISE mass determinations,
(i.e. the average ratio M$_{\rm ATLAS3D}$/M$_{\rm 3.4\mu m}$ is 0.86 not 1.) in the sense that the WISE 
determinations are higher. The fact that the magnitude of this offset is very similar for both W1 and W2 
indicates that this could be related to the method used to determine the total luminosity in the WISE bands, 
and that we have systematically overestimated the total luminosity, or, alternatively, that the stellar population
parameters (principally the ages) determined by the SAURON survey are differ systematically from those
used by the ATLAS3D survey in their analysis.

\item A third observation is that the scatter is reduced compared to that found for the fixed M$_\star$/L case. 
When it is considered that the ATLAS3D stellar masses contribute a significant fraction of the total scatter
the magnitude of the decrease is even more remarkable. 

\item Finally, and most intriguingly, there is
evidence that the relation is no longer one-to-one, in the sense that the WISE derived masses 
are increasingly over-massive compared to the ATLAS3D ones for higher mass galaxies. We leave
a more detailed investigation of this point to a forthcoming paper.

\end{enumerate}

\section{Discussion}

We have used NIR WISE photometry of a diverse sample of old stellar systems to probe the behavior
of these stellar systems as a function of the full metallicity range displayed by star clusters and galaxies.
Using this sample we have confirmed that the latest generation of stellar population synthesis models 
(in particular the Padova group models of \citealt{Bressan12,Bressan13}) are capable of accurately 
reproducing the colors and luminosities of the 3.4 and 4.6$\mu$m bands for old stellar systems. 

Having confirmed that the SPS models accurately reproduce the NIR photometry of real stellar systems, 
we then made use of the models to derive mass-to-light ratios for single burst stellar populations as a 
function of metallicity and age. As found by \citet{Meidt14a} for the IRAC 3.6$\mu$m band this procedure 
demonstrated that the WISE W1 and W2 M$_\star$/L ratios are almost insensitive to metallicity for [Fe/H] $>$ --1 
dex. Furthermore, by comparison with the stellar masses derived by the ATLAS3D survey we confirmed
that the use of a single fixed M$_\star$/L in the NIR for galaxies older than 3 Gyr can produce remarkably accurate
stellar mass estimates. 

Having confirmed that modern SPS models can be used to accurately predict stellar masses for older 
stellar populations several further steps can be envisaged to make the technique more widely applicable.
In order to extend this technique to younger stellar populations several confusing effects will have to be
integrated into the models or the analysis procedure. The most pressing is to include the effect of evolved 
stellar phases such as TP-AGB stars, which can contribute significantly to the total luminosity in the NIR 
for ages $<$ 3 Gyr, even in cases where the galaxy (though not the AGB stars) itself may be relatively 
dust-free. Implementation of this change is already underway by the Padova group and other SPS modellers. 

A second necessary step will be to determine and remove the effect of non-stellar emission in the NIR bands, 
particularly in the 3.6$\mu$m band, where a PAH emission feature is present. Removal of non-stellar emission 
could potentially be achieved using several approaches, one would be to use independent component analysis 
such as was used for the S$^{\rm 4}$G sample by \citet{Meidt12a} and Querejeta et al. (2014). An alternative 
approach that could be more successful for poorly resolved (or entirely un-resolved) sources might be to make 
use of the W3 band as a probe of the non-stellar emission and to use this to correct the W1 photometry.

\section{Conclusions}

We have presented the first large sample size examination of the dust-free NIR photometry of older stellar
populations as a function of the full metallicity range displayed by globular clusters and galaxies. Our main 
conclusions are:

\begin{enumerate}

\item Contrary to the predictions of the majority of SSP models the W1-W2 colors of stellar populations 
become systematically \emph{bluer} with increasing metallicity.

\item Using SPS models which fit the observed color-metallicity relations, and hence accurately predict both
the 3.4 and 4.6$\mu$m fluxes we derive mass-to-light ratios for both the WISE W1 and W2 filters. In doing
this we determined that for the range of metallicities encountered in massive galaxies the M$_\star$/L in the NIR is 
relatively insensitive to metallicity, and in the case of age, varies by a factor of 2 between 3 and 10 Gyr.

\item By comparison of our WISE-derived stellar masses with a SAURON+ATLAS3D sample of 
early-type galaxies we confirm the finding of \citet{Meidt14a} that a single fixed M$_\star$/L at 3.4$\mu$m can 
produce stellar masses with uncertainties of only 0.1 dex.

\item We find that when including additional age information the accuracy of the derived stellar masses can 
be improved by a further 0.02 dex for dust free stellar populations older than 3 Gyr.

\end{enumerate}

\section{Acknowledgements}

We would like to thank Benjamin R\"ock for useful discussions regarding modelling the NIR emissions
of old stellar populations.

We would like to thank Agnieszka Ry\'s for providing the measured stellar population parameters for
her sample of dwarf galaxies ahead of publication.

This publication makes use of data products from the Wide-field Infrared Survey Explorer, which is a joint 
project of the University of California, Los Angeles, and the Jet Propulsion Laboratory/California Institute 
of Technology, and NEOWISE, which is a project of the Jet Propulsion Laboratory/California Institute of 
Technology. WISE and NEOWISE are funded by the National Aeronautics and Space Administration.

We acknowledge financial support to the DAGAL network from the People Programme (Marie Curie Actions) 
of the European UnionÕs Seventh Framework Programme FP7/2007- 2013/ under REA grant agreement number 
PITN-GA-2011-289313

We acknowledge the usage of the HyperLeda database (http://leda.univ-lyon1.fr).

This research has made use of the NASA/IPAC Extragalactic Database (NED) which is operated 
by the Jet Propulsion Laboratory, California Institute of Technology, under contract with the 
National Aeronautics and Space Administration. 
 
Funding for SDSS-III has been provided by the Alfred P. Sloan Foundation, the Participating Institutions, 
the National Science Foundation, and the U.S. Department of Energy Office of Science. The SDSS-III 
web site is http://www.sdss3.org/.

SDSS-III is managed by the Astrophysical Research Consortium for the Participating Institutions of the 
SDSS-III Collaboration including the University of Arizona, the Brazilian Participation Group, Brookhaven 
National Laboratory, Carnegie Mellon University, University of Florida, the French Participation Group, the 
German Participation Group, Harvard University, the Instituto de Astrofisica de Canarias, the Michigan 
State/Notre Dame/JINA Participation Group, Johns Hopkins University, Lawrence Berkeley National Laboratory, 
Max Planck Institute for Astrophysics, Max Planck Institute for Extraterrestrial Physics, New Mexico State 
University, New York University, Ohio State University, Pennsylvania State University, University of Portsmouth, 
Princeton University, the Spanish Participation Group, University of Tokyo, University of Utah, Vanderbilt 
University, University of Virginia, University of Washington, and Yale University.

\section{Appendix}

\begin{table}
\begin{center}
\begin{tabular}{llll} \hline
Name				& [Fe/H]		& W1 - W2									 \\
					& [dex]		& [mag]										\\
\hline
\multicolumn{3}{l}{\bf Milky Way Globular Clusters}& \\
\\
NGC~1261 &         -1.27 & -0.008 $\pm$ 0.009 \\ 
Pal2    &         -1.42 &  0.025 $\pm$ 0.012 \\ 
NGC~1851 &         -1.18 & -0.026 $\pm$ 0.009 \\ 
NGC~1904 &         -1.60 &  0.005 $\pm$ 0.009 \\ 
NGC~2419 &         -2.15 & -0.005 $\pm$ 0.019 \\ 
NGC~3201 &         -1.59 & -0.011 $\pm$ 0.012 \\ 
NGC~4590 &         -2.23 &  0.022 $\pm$ 0.010 \\ 
NGC~4833 &         -1.85 &  0.020 $\pm$ 0.012 \\ 
NGC~5024 &         -2.10 & -0.009 $\pm$ 0.010 \\ 
$\omega$Cen      &  -1.53 &  0.019 $\pm$ 0.010 \\ 
NGC~5272 &         -1.50 & -0.008 $\pm$ 0.009 \\ 
NGC~5286 &         -1.69 & -0.015 $\pm$ 0.009 \\ 
NGC~5634 &         -1.88 &  0.005 $\pm$ 0.009 \\ 
NGC~5694 &         -1.98 & -0.008 $\pm$ 0.036 \\ 
IC4499 &         -1.53 & -0.025 $\pm$ 0.026 \\ 
NGC~5824 &         -1.91 &  0.023 $\pm$ 0.009 \\ 
NGC~5897 &         -1.90 &  0.022 $\pm$ 0.015 \\ 
NGC~5904 &         -1.29 & -0.018 $\pm$ 0.019 \\ 
NGC~5946 &         -1.29 & -0.001 $\pm$ 0.015 \\ 
NGC~5986 &         -1.59 & -0.004 $\pm$ 0.009 \\ 
NGC~6093 &         -1.75 &  0.004 $\pm$ 0.009 \\ 
NGC~6139 &         -1.65 &  0.006 $\pm$ 0.010 \\ 
NGC~6171 &         -1.02 &  0.028 $\pm$ 0.010 \\ 
NGC~6205 &         -1.53 & -0.006 $\pm$ 0.009 \\ 
NGC~6229 &         -1.47 & -0.034 $\pm$ 0.009 \\ 
NGC~6218 &         -1.37 &  0.008 $\pm$ 0.009 \\ 
NGC~6254 &         -1.56 &  0.037 $\pm$ 0.009 \\ 
NGC~6273 &         -1.74 & -0.003 $\pm$ 0.011 \\ 
NGC~6284 &         -1.26 & -0.044 $\pm$ 0.011 \\ 
NGC~6287 &         -2.10 &  0.011 $\pm$ 0.011 \\ 
NGC~6293 &         -1.99 &  0.017 $\pm$ 0.019 \\ 
NGC~6316 &         -0.45 & -0.025 $\pm$ 0.021 \\ 
NGC~6341 &         -2.31 &  0.008 $\pm$ 0.009 \\ 
NGC~6402 &         -1.28 & -0.006 $\pm$ 0.009 \\ 
NGC~6496 &         -0.46 & -0.065 $\pm$ 0.018 \\ 
NGC~6517 &         -1.23 &  0.044 $\pm$ 0.012 \\ 
NGC~6539 &         -0.63 &  0.008 $\pm$ 0.014 \\ 
NGC~6541 &         -1.81 &  0.010 $\pm$ 0.009 \\ 
NGC~6569 &         -0.76 & -0.018 $\pm$ 0.020 \\ 
NGC~6584 &         -1.50 & -0.026 $\pm$ 0.011 \\ 
NGC~6624 &         -0.44 & -0.010 $\pm$ 0.012 \\ 
NGC~6638 &         -0.95 & -0.001 $\pm$ 0.013 \\ 
NGC~6681 &         -1.62 & -0.023 $\pm$ 0.010 \\ 
NGC~6712 &         -1.02 & -0.050 $\pm$ 0.028 \\ 
NGC~6723 &         -1.10 & -0.025 $\pm$ 0.017 \\ 
NGC~6779 &         -1.98 & -0.002 $\pm$ 0.010 \\ 
NGC~6864 &         -1.29 &  0.017 $\pm$ 0.009 \\ 
NGC~6934 &         -1.47 & -0.020 $\pm$ 0.009 \\ 
NGC~6981 &         -1.42 & -0.016 $\pm$ 0.011 \\ 
NGC~7006 &         -1.52 & -0.033 $\pm$ 0.010 \\ 
NGC~7078 &         -2.37 &  0.006 $\pm$ 0.009 \\ 
NGC~7089 &         -1.65 & -0.004 $\pm$ 0.009 \\ 
NGC~7099 &         -2.27 &  0.004 $\pm$ 0.009 \\ 
\hline
\end{tabular}
\caption[Catalog]{Cataloged metallicity and WISE W1-W2 color as measured using the approach described in
Section 2.2 and plotted in Figure~\ref{fig:FeH_color}.}
\end{center}
\label{tab:catalog}
\end{table}

\begin{table}
\begin{center}
\begin{tabular}{llll} \hline
Name				& [Fe/H]		& W1 - W2									 \\
					& [dex]		& [mag]										\\
\hline
\\
\multicolumn{3}{l}{\bf LMC and SMC Globular Clusters}& \\
\\
HODGE11 &       -2.06 & -0.014 $\pm$ 0.020 \\ 
NGC~1466 &       -2.17 & -0.025 $\pm$ 0.016 \\ 
NGC~1754 &       -1.54 & -0.034 $\pm$ 0.022 \\ 
NGC~1786 &       -1.87 & -0.038 $\pm$ 0.008 \\ 
NGC~1841 &       -2.11 & -0.002 $\pm$ 0.030 \\ 
NGC~1898 &       -1.37 & -0.036 $\pm$ 0.015 \\ 
NGC~2121 &       -0.61 & -0.004 $\pm$ 0.032 \\ 
NGC~2173 &       -0.24 & -0.087 $\pm$ 0.014 \\ 
NGC~2210 &       -1.97 & -0.032 $\pm$ 0.012 \\ 
KRON3 &       -1.16 & -0.032 $\pm$ 0.035 \\ 
NGC~121 &       -1.71 & -0.013 $\pm$ 0.012 \\ 
NGC~339 &       -1.50 & -0.071 $\pm$ 0.043 \\ 
NGC~361 &       -1.45 & -0.064 $\pm$ 0.024 \\ 
\\
\multicolumn{3}{l}{\bf M31 Globular Clusters}& \\
\\
M31-B001 &         -0.70 & -0.021 $\pm$ 0.043 \\ 
M31-B004 &         -0.70 & -0.077 $\pm$ 0.046 \\ 
M31-B005 &         -0.70 & -0.058 $\pm$ 0.037 \\ 
M31-B006 &         -0.50 & -0.060 $\pm$ 0.033 \\ 
M31-B008 &         -0.80 & -0.092 $\pm$ 0.045 \\ 
M31-B012 &         -1.70 & -0.033 $\pm$ 0.033 \\ 
M31-B013 &         -0.50 & -0.092 $\pm$ 0.050 \\ 
M31-B017 &         -0.82 & -0.040 $\pm$ 0.035 \\ 
M31-B019 &         -0.80 & -0.043 $\pm$ 0.033 \\ 
M31-B020 &         -0.90 & -0.060 $\pm$ 0.033 \\ 
M31-B023 &         -0.70 & -0.043 $\pm$ 0.030 \\ 
M31-B024 &         -0.60 & -0.085 $\pm$ 0.045 \\ 
M31-B027 &         -1.30 & -0.045 $\pm$ 0.048 \\ 
M31-B034 &         -0.60 & -0.078 $\pm$ 0.042 \\ 
M31-B037 &         -0.80 & -0.032 $\pm$ 0.035 \\ 
M31-B039 &         -0.80 &  0.021 $\pm$ 0.040 \\ 
M31-B045 &         -0.90 & -0.083 $\pm$ 0.037 \\ 
M31-B050 &         -0.80 & -0.011 $\pm$ 0.050 \\ 
M31-B051 &         -0.80 & -0.102 $\pm$ 0.039 \\ 
M31-B058 &         -1.10 & -0.037 $\pm$ 0.035 \\ 
M31-B061 &         -0.70 & -0.019 $\pm$ 0.042 \\ 
M31-B063 &         -0.80 & -0.020 $\pm$ 0.034 \\ 
M31-B068 &         -0.20 & -0.070 $\pm$ 0.047 \\ 
M31-B074 &         -1.50 & -0.044 $\pm$ 0.049 \\ 
M31-B082 &         -0.70 & -0.064 $\pm$ 0.035 \\ 
M31-B088 &         -1.80 & -0.026 $\pm$ 0.034 \\ 
M31-B094 &         -0.40 & -0.071 $\pm$ 0.035 \\ 
M31-B110 &         -0.70 & -0.050 $\pm$ 0.042 \\ 
M31-B116 &         -0.60 & -0.051 $\pm$ 0.040 \\ 
M31-B135 &         -1.46 & -0.045 $\pm$ 0.047 \\ 
M31-B158 &         -0.74 & -0.094 $\pm$ 0.033 \\ 
M31-B163 &         -0.29 & -0.038 $\pm$ 0.043 \\ 
M31-B174 &         -1.00 & -0.061 $\pm$ 0.040 \\ 
M31-B182 &         -1.00 & -0.102 $\pm$ 0.037 \\ 
M31-B183 &         -0.50 & -0.040 $\pm$ 0.040 \\ 
M31-B193 &         -0.10 & -0.055 $\pm$ 0.041 \\ 
M31-B212 &         -1.70 & -0.053 $\pm$ 0.038 \\ 
M31-B218 &         -0.80 & -0.047 $\pm$ 0.038 \\ 
M31-B219 &         -0.01 & -0.063 $\pm$ 0.040 \\ 
M31-B225 &         -0.44 & -0.028 $\pm$ 0.033 \\ 
M31-B232 &         -1.90 & -0.016 $\pm$ 0.042 \\ 
M31-B233 &         -1.10 & -0.052 $\pm$ 0.044 \\ 
M31-B238 &         -0.57 & -0.004 $\pm$ 0.043 \\ 
M31-B240 &         -1.50 & -0.047 $\pm$ 0.035 \\ 
M31-B301 &         -1.19 & -0.010 $\pm$ 0.049 \\ 
M31-B306 &         -1.10 & -0.047 $\pm$ 0.033 \\ 
M31-B311 &         -1.90 & -0.050 $\pm$ 0.035 \\ 
M31-B312 &         -1.20 & -0.041 $\pm$ 0.035 \\ 
M31-B313 &         -0.83 &  0.006 $\pm$ 0.039 \\ 
M31-B344 &         -1.00 & -0.053 $\pm$ 0.038 \\ 
M31-B348 &         -1.38 & -0.019 $\pm$ 0.044 \\ 
\hline
\end{tabular}
\caption[Catalog]{Table 1 cont.}
\end{center}
\label{tab:catalog2}
\end{table}

\begin{table}
\begin{center}
\begin{tabular}{llll} \hline
Name				& [Fe/H]		& W1 - W2									 \\
					& [dex]		& [mag]										\\
\hline
\\
\multicolumn{3}{l}{\bf M31 Globular Clusters cont.}& \\
\\
M31-B352 &         -1.50 & -0.032 $\pm$ 0.050 \\ 
M31-B373 &         -0.50 & -0.032 $\pm$ 0.040 \\ 
M31-B379 &         -0.40 & -0.021 $\pm$ 0.035 \\ 
M31-B381 &         -1.10 & -0.022 $\pm$ 0.037 \\ 
M31-B383 &         -0.57 & -0.034 $\pm$ 0.034 \\ 
M31-B384 &         -0.70 & -0.014 $\pm$ 0.035 \\ 
M31-B386 &         -1.10 & -0.001 $\pm$ 0.036 \\ 
M31-B397 &         -1.20 &  0.053 $\pm$ 0.043 \\ 
M31-B405 &         -1.20 & -0.046 $\pm$ 0.033 \\ 

\\
\multicolumn{3}{l}{\bf SAURON Galaxies}& \\
\\
NGC~0474 &       -0.25 & -0.032 $\pm$ 0.014 \\ 
NGC~0524 &       -0.16 & -0.032 $\pm$ 0.006 \\ 
NGC~0821 &       -0.36 & -0.027 $\pm$ 0.009 \\ 
NGC~1023 &       -0.27 & -0.040 $\pm$ 0.004 \\ 
NGC~2549 &       -0.12 & -0.063 $\pm$ 0.009 \\ 
NGC~2685 &       -0.45 & -0.034 $\pm$ 0.011 \\ 
NGC~2695 &       -0.54 & -0.072 $\pm$ 0.013 \\ 
NGC~2768 &       -0.45 & -0.030 $\pm$ 0.006 \\ 
NGC~3377 &       -0.42 & -0.036 $\pm$ 0.007 \\ 
NGC~3379 &       -0.38 & -0.048 $\pm$ 0.004 \\ 
NGC~3384 &       -0.16 & -0.047 $\pm$ 0.006 \\ 
NGC~3414 &       -0.47 & -0.052 $\pm$ 0.009 \\ 
NGC~3489 &       -0.20 & -0.023 $\pm$ 0.007 \\ 
NGC~3608 &       -0.34 & -0.056 $\pm$ 0.009 \\ 
NGC~4150 &       -0.27 &  0.008 $\pm$ 0.014 \\ 
NGC~4270 &       -0.33 & -0.062 $\pm$ 0.018 \\ 
NGC~4278 &       -0.46 & -0.059 $\pm$ 0.006 \\ 
NGC~4374 &       -0.44 & -0.054 $\pm$ 0.004 \\ 
NGC~4382 &       -0.31 & -0.029 $\pm$ 0.004 \\ 
NGC~4387 &       -0.41 & -0.084 $\pm$ 0.017 \\ 
NGC~4458 &       -0.63 & -0.049 $\pm$ 0.019 \\ 
NGC~4459 &       -0.32 & -0.035 $\pm$ 0.006 \\ 
NGC~4473 &       -0.34 & -0.065 $\pm$ 0.006 \\ 
NGC~4477 &       -0.37 & -0.053 $\pm$ 0.007 \\ 
NGC~4486 &       -0.44 & -0.029 $\pm$ 0.003 \\ 
NGC~4526 &       -0.27 & -0.016 $\pm$ 0.005 \\ 
NGC~4546 &       -0.42 & -0.077 $\pm$ 0.007 \\ 
NGC~4552 &       -0.21 & -0.072 $\pm$ 0.005 \\ 
NGC~4564 &       -0.29 & -0.075 $\pm$ 0.010 \\ 
NGC~4621 &       -0.41 & -0.061 $\pm$ 0.005 \\ 
NGC~5198 &       -0.36 & -0.061 $\pm$ 0.014 \\ 
NGC~5813 &       -0.41 & -0.030 $\pm$ 0.008 \\ 
NGC~5831 &       -0.23 & -0.042 $\pm$ 0.011 \\ 
NGC~5838 &       -0.26 & -0.049 $\pm$ 0.008 \\ 
NGC~5846 &       -0.38 & -0.045 $\pm$ 0.006 \\ 
NGC~5982 &       -0.20 & -0.055 $\pm$ 0.009 \\ 
NGC~7457 &       -0.23 &  0.000 $\pm$ 0.010 \\ 

\hline
\end{tabular}
\caption[Catalog]{Table 1 cont.}
\end{center}
\label{tab:catalog3}
\end{table}

\begin{table}
\begin{center}
\begin{tabular}{llll} \hline
Name				& [Fe/H]		& W1 - W2									 \\
					& [dex]		& [mag]										\\
\hline
\\
\multicolumn{3}{l}{\bf Other Dwarf and Giant ETGs}& \\
\\
VCC0308 &         -0.97 & -0.054 $\pm$ 0.072 \\ 
VCC0490 &         -0.35 & -0.031 $\pm$ 0.081 \\ 
VCC0523 &         -0.81 & -0.051 $\pm$ 0.044 \\ 
VCC0543 &         -0.41 & -0.037 $\pm$ 0.076 \\ 
VCC0634 &         -0.30 & -0.039 $\pm$ 0.070 \\ 
VCC0856 &         -0.43 & -0.096 $\pm$ 0.085 \\ 
VCC0929 &         -0.79 & -0.057 $\pm$ 0.040 \\ 
VCC1010 &         -0.33 & -0.079 $\pm$ 0.041 \\ 
VCC1036 &         -0.75 & -0.048 $\pm$ 0.048 \\ 
VCC1087 &         -1.10 & -0.064 $\pm$ 0.060 \\ 
VCC1261 &         -0.98 & -0.036 $\pm$ 0.048 \\ 
VCC1422 &         -0.48 & -0.084 $\pm$ 0.057 \\ 
VCC1861 &         -0.83 & -0.030 $\pm$ 0.069 \\ 
VCC1910 &         -0.13 & -0.088 $\pm$ 0.053 \\ 
VCC1912 &         -0.48 & -0.065 $\pm$ 0.068 \\ 
VCC2019 &         -0.28 & -0.102 $\pm$ 0.098 \\ 
NGC~3073 &         -0.87 &  0.006 $\pm$ 0.040 \\ 
ID0650 &         -1.00 & -0.054 $\pm$ 0.076 \\ 
VCC0731 &         -0.14 & -0.085 $\pm$ 0.006 \\ 
VCC0828 &         -0.78 & -0.065 $\pm$ 0.015 \\ 
VCC1146 &         -1.17 & -0.063 $\pm$ 0.022 \\ 
VCC1279 &         -0.31 & -0.077 $\pm$ 0.011 \\ 
VCC1630 &         -0.54 & -0.066 $\pm$ 0.017 \\ 
VCC1903 &         -0.23 & -0.076 $\pm$ 0.006 \\ 
FCC021 &          0.07 & -0.022 $\pm$ 0.003 \\ 
FCC167 &         -0.21 & -0.053 $\pm$ 0.008 \\ 
FCC043 &         -0.50 & -0.041 $\pm$ 0.045 \\ 
FCC136 &         -0.65 & -0.044 $\pm$ 0.084 \\ 
FCC335 &         -0.52 & -0.023 $\pm$ 0.054 \\ 
NGC~205 &         -2.04 &  0.003 $\pm$ 0.005 \\ 
NGC~404 &         -0.38 & -0.003 $\pm$ 0.007 \\ 
NGC~0720 &          0.25 & -0.063 $\pm$ 0.006 \\ 
NGC~1045 &          0.33 & -0.078 $\pm$ 0.016 \\ 
NGC~1132 &         -0.45 & -0.055 $\pm$ 0.021 \\ 
NGC~1407 &          0.29 & -0.050 $\pm$ 0.006 \\ 
NGC~1453 &          0.06 & -0.069 $\pm$ 0.011 \\ 
NGC~1700 &          0.14 & -0.055 $\pm$ 0.010 \\ 
NGC~2128 &         -0.10 & -0.008 $\pm$ 0.013 \\ 
NGC~2513 &          0.26 & -0.068 $\pm$ 0.015 \\ 
NGC~2911 &          0.26 & -0.035 $\pm$ 0.014 \\ 
NGC~3091 &          0.39 & -0.062 $\pm$ 0.010 \\ 
NGC~3226 &         -0.52 & -0.016 $\pm$ 0.012 \\ 
NGC~3607 &          0.30 & -0.062 $\pm$ 0.006 \\ 
NGC~3613 &          0.07 & -0.058 $\pm$ 0.010 \\ 
NGC~3640 &          0.21 & -0.048 $\pm$ 0.008 \\ 
NGC~3923 &          0.33 & -0.052 $\pm$ 0.005 \\ 
NGC~5812 &          0.24 & -0.097 $\pm$ 0.012 \\ 
NGC~5846A &        -0.36 & -0.053 $\pm$ 0.019 \\ 
NGC~7302 &          0.16 & -0.069 $\pm$ 0.017 \\ 
NGC~7626 &         -0.12 & -0.052 $\pm$ 0.011 \\ 

\hline
\end{tabular}
\caption[Catalog]{Table 1 cont.}
\end{center}
\label{tab:catalog4}
\end{table}

\bibliographystyle{apj}
\bibliography{references}
\label{lastpage}

\end{document}